# A Prototype Scintillator Real-Time Beam Monitor for Ultra-high Dose Rate Radiotherapy

Daniel S. Levin[1],  Peter S. Friedman[2]

Claudio Ferretti[1], Nicholas Ristow[1], Monica Tecchio[1]

Dale W. Litzenberg[3]

Vladimir Bashkirov[4], Reinhard Schulte[4]

1.  Department of Physics, University of Michigan, Ann Arbor, MI, 48109, USA
2.  Integrated Sensors LLC, Palm Beach Gardens, FL 33418, USA
3.  Department of Radiation Oncology, University of Michigan, Ann Arbor, MI, 48109, USA
4.  Division of Biomedical Engineering Sciences, Loma Linda University School of Medicine, Loma Linda, CA 92350, USA

Contact: Daniel Levin, Department of Physics, University of Michigan. dslevin@umich.edu

## ABSTRACT

### Background

FLASH Radiotherapy (RT) is an emergent cancer radiotherapy modality where an entire therapeutic dose is delivered at more than 1000 times higher dose rate than conventional RT. For clinical trials to be conducted safely, a precise and fast beam monitor that can generate out-of-tolerance beam interrupts is required. This paper describes the overall concept and provides results from a prototype ultra-fast, scintillator-based beam monitor for both proton and electron beam FLASH applications.

### Purpose

A FLASH Beam Scintillator Monitor (FBSM) is being developed that employs a novel proprietary scintillator material. The FBSM has capabilities that conventional RT detector technologies are unable to simultaneously provide: 1) large area coverage; 2) a low mass profile; 3) a linear response over a broad dynamic range; 4) radiation hardness; 5) real-time analysis to provide an IEC-compliant fast beam-interrupt signal based on true two-dimensional beam imaging, radiation dosimetry and excellent spatial resolution.

### Methods

The FBSM uses a proprietary low mass, less than 0.5 mm water equivalent, non-hygroscopic, radiation tolerant scintillator material (designated HM: hybrid material) that is viewed by high frame rate CMOS cameras. Folded optics using mirrors enable a thin monitor profile of ~10 cm. A field programmable gate array (FPGA) data acquisition system (DAQ) generates real-time analysis on a time scale appropriate to the FLASH RT beam modality: 100-1000 Hz for pulsed electrons and 10-20 kHz for quasi-continuous scanning proton pencil beams. An ion beam monitor served as the initial development platform for this work and was tested in low energy heavy-ion beams ($^{86}Kr^{+26}$ and protons). A prototype FBSM was fabricated and then tested in various radiation beams that included FLASH level dose per pulse electron beams, and a hospital radiotherapy clinic with electron beams.

### Results

Results presented in this report include image quality, response linearity, radiation hardness, spatial resolution, and real-time data processing. The HM scintillator was found to be highly radiation damage resistant. It exhibited a small 0.025%/kGy signal decrease from a 216 kGy cumulative dose resulting from continuous exposure for 15 minutes at a FLASH compatible dose rate of 237 Gy/s. Measurements of the signal amplitude vs beam fluence demonstrate linear response of the FBSM at FLASH compatible dose rates of > 40 Gy/s. Comparison with commercial Gafchromic film indicates that the FBSM produces a high resolution 2D beam image and can reproduce a nearly identical beam profile, including primary beam tails. The spatial resolution was measured at 35-40 μm. Tests of the firmware beta version show successful operation at 20,000 Hz frame rate or 50 μs/frame, where the real-time analysis of the beam parameters is achieved in less than 1 μs.

## Conclusions

The FBSM is designed to provide real-time beam profile monitoring over a large active area without significantly degrading the beam quality. A prototype device has been staged in particle beams at currents of single particles up to FLASH level dose rates, using both continuous ion beams and pulsed electron beams. Using a novel scintillator, beam profiling has been demonstrated for currents extending from single particles to 10 nA currents. Radiation damage is minimal and even under FLASH conditions would require ≥ 50 kGy of accumulated exposure in a single spot to result in a 1% decrease in signal output. Beam imaging is comparable to radiochromic films, and provides immediate images without hours of processing. Real-time data processing, taking less than 50 μs (combined data transfer and analysis times), has been implemented in firmware for 20 kHz frame rates for continuous proton beams.

## 1 INTRODUCTION

FLASH radiotherapy (RT) is an emerging modality for cancer treatment in which normal tissue toxicity is reduced using ultra-high dose rates exceeding 40 Gy/s. Several pre-clinical animal trials used pulsed electron beams [1, 2, 3, 4] and scanned pencil beam protons on mice [5, 6]. The first human clinical trial (FAST-01) was done with a 250-MeV proton transmission beam [7]. In electron-FLASH [8, 9, 10, 11, 12], a linac generates a sequence of short 0.5-5 μs pulses at a repetition rate in the range of 100-1000 Hz, and very high electron energies (VHEE) can extend upwards from a minimum of 100 MeV. Proton-FLASH consists of passively scattered or scanned pencil beams typically from isochronous cyclotrons, which have a quasi-continuous structure consisting of ~2 ns duration bunches at a frequency of roughly 70-130 MHz, or synchrocyclotrons, which produce beams of microsecond pulses every few milliseconds. In any of these FLASH RT delivery modes, the time-averaged dose rate can be at least two or even three orders of magnitude higher than in conventional RT. Notably, the instantaneous dose rate is orders of magnitude higher still, extending to more than $10^9$ Gy/second [13].

Precise and fast real-time beam monitors are required for future patient treatments to be conducted effectively and safely. The quasi-continuous versus pulsed nature of these beams dictates some aspects of the monitor components, as described below. Instrumentation for conventional RT dose measurement includes ionization chambers, and passive radiochromic films that provide detailed and precise dosimetry information minutes to hours after exposure. Conventional and FLASH real-time beam monitoring is typically implemented with strip ionization chambers or inductive beam current transformers, for example, the unit made by Bergoz (Saint Genis Pouilly, France) [14], which is used in the Mobetron (IntraOp, Sunnyvale, CA, USA) medical linac. However, these beam monitoring instruments are not in general optimized for FLASH beam applications as they may saturate or lose charge collection efficiency at high rates, can have slow response times, have insufficient spatial resolution, include assumptions about the beam profile [15], have limited area coverage, or introduce excessive mass in the beam path.

To the best of our knowledge, no single technology provides the combined attributes of large area coverage, a low mass profile, small longitudinal footprint, large dynamic range with a linear response, high radiation hardness, true two-dimensional imaging with high spatial resolution, and most critically for FLASH operation, the capability to generate a real-time beam-interrupt if the beam current or beam profile deviates from a prescribed irradiation plan. In the context of this application, the real-time or response time interval depends on the type of beam applied. As noted above, electron beams have pulse repetition rates of 100-1000 Hz so that FBSM response time to monitor these pulses can range from 1-10 ms. Scanning proton beams are effectively continuous, and the response time should be as short as possible, i.e. 50 µs as outlined in this paper. The primary determinants of these times are the camera frame rate capability and the speed of data flow and analysis, discussed in detail in the following section. This paper describes a FLASH therapy compatible Beam Scintillator Monitor (FBSM) design with these capabilities. It employs a proprietary scintillator in a novel application viewed by fast and ultra-fast CMOS based cameras. The data acquisition system (DAQ) and beam analysis are embedded in a field programmable gate array (FPGA) whose firmware is written in Verilog. The first experimental results are reported for a prototype device and its associated components that have been tested with radiation sources, using pulsed electron beams at the Notre Dame Radiation Laboratory (NDRL), and at the University of Michigan Hospital Radiation Oncology (UMH). Related beam tests with an ion beam monitor that served as a development platform for the current effort were conducted at the DOE Facility for Rare Isotope Beams (FRIB) [16].

## 1.1 Other Technologies

The promise of the emergence of the FLASH modality has prompted development of several beam monitoring technologies that might be deployed in experimental work and eventually in clinical trials and conventional therapeutic usage. Several research programs are exploring FLASH-compatible beam monitors [15, 17, 18, 19, 20, 21, 22, 23, 24, 25].

Beam current transformer monitors [14, 17, 19, 20, 21] gauge the beam intensity without introduction of a mass layer. They have been tested in a pulsed electron beam where the pulse rate is generally 100 Hz or less. Operation at the much higher rates necessary for scanning nearly continuous proton beams has not been demonstrated to our knowledge. The drawback of beam current transformers is they only measure the flux of electrons and do not measure beam shape and spatial distribution. Thus, they must be coupled with ionization chambers [23] or other imaging detectors to obtain profile information.

Other approaches use small inorganic scintillating crystals or organic scintillators coupled to fibers [18, 26] and read out by photodetectors. These have demonstrated potentially high-rate capabilities with 20 µs time resolution for use in proton beams. Beam profile imaging however requires large planar arrays assembled from multiple crystals, with a commensurate number of readout channels. The relatively large thickness of these scintillators, which can exceed 1 mm water equivalent thickness (WET), can also have a degrading effect on the beam. In addition, an array of multiple crystals will always have dead detection regions between adjacent crystals.

In addition to the above, there are also new efforts to develop fast time-resolved beam imaging dosimeters that use fast scintillators. Such detectors, when placed in a proton beam, produce a scintillation light on a pulse-by-pulse basis [22]. Another imaging system uses a CMOS camera to view a scintillator, but at a large distance, and was tested in a scanning pencil proton beam [24, 27]. This is discussed later.

## 2 METHODS

In this section we describe the monitor in Sections 2.1-2.4 and a sequence of experiments conducted with the prototype FBSM and associated components. These experiments include optical setups on the laboratory testbench that use a collimated beta source (Section 2.6), staging of the prototype in conventional electron RT beams at UMH (Section 2.7), and an ultra-high dose rate, FLASH-compatible electron beam at NDRL (Section 2.8). For completeness, we review experiments at FRIB using major FBSM components (Section 2.9).

### 2.1 Monitor Design

The FBSM employs a proprietary low mass scintillator through which the beam passes with minimal energy loss. This paper describes a 1st generation prototype FBSM that accommodates a 15 cm × 15 cm sensitive area viewed by a single camera. A 2nd generation FBSM currently under development hosts a 15 cm × 23 cm sensitive area and employs two cameras with overlapping image fields of view. All optical components are mounted in a light-tight enclosure along with blackout materials including flocked paper and black carbon nanofiber paints to eliminate ambient background and reduce internal reflections. All critical design elements are manifested in the prototype FBSM shown in Figure 1.

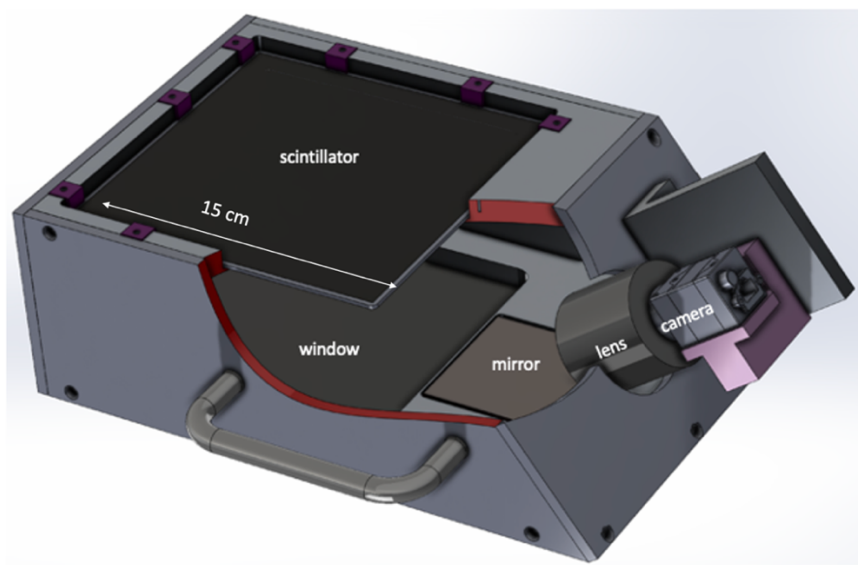

Figure 1: Engineering drawing of prototype FBSM. The camera is shielded by a graphite-Pb radiation shield (not shown). The enclosure is light tight to minimize ambient light background.

The total longitudinal mass profile for the FBSM, including the scintillator and the entrance/exit windows was determined using the GEANT4 program [28] and the PSTAR and ESTAR databases [29]. For protons in the energy range 170-240 MeV the thickness is ~0.57 mm WET and the maximum energy loss is ~0.32 MeV or < 0.2% of the particle energy. The corresponding maximum loss in proton range is 0.78 mm.  For a 100 MeV VHEE beam, the WET is ~0.5 mm. For a 6-16 MeV electron beam, the maximum energy loss at 16 MeV is ~0.16 MeV, or 1% and corresponding range loss is 0.7 mm WET. Overall, the low mass ensures that the monitor is transmissive with minimal particle energy loss and low multiple scattering.

The active area of the transmissive scintillator is viewed by CMOS-sensor camera(s), which operate in either a synchronous triggered or an asynchronous, untriggered, quasi-continuous global shutter mode. The triggered mode is intended for electron beams that can have pulse repetition rates up to ~1000 Hz. The untriggered mode is intended for quasi-continuous proton beams. For the pulsed electron beams, the FBSM incorporates a specific class of low noise cameras, denoted here as CamE, characterized by nearly 2 megapixels (MP) at full resolution, better than 70 dB dynamic range and readout noise of ~2.5 electrons ($e^-$).  Nominal high radiation peak signals extend to several thousand $e^-$, depending on the specific dose rate and exposure duration.  A continuous DAQ mode is in development for scanning proton beams at frame rates extending to 10-20 kHz, with first results reported in Section 3.1.3.  In this type of operation, a synchronization signal from the beam initiates the camera's data acquisition, which would then run continuously at a high frame rate for the duration of the scan, with negligible dead-time. The proton beam camera, denoted as CamP, has a 1 MP sensor with ~60 dB dynamic range and an RMS readout noise of < 15 e. Operation at these high frame rates is achieved by limiting the active sensor area to be read out, or Region of Interest (ROI). This reduces the number of pixels read out with a corresponding reduction in the spatial resolution. At the full frame rate of 20 kHz and 50 $\mu$s time resolution, the ROI of CamP is approximately 45,000 pixels, but still results in a spatial resolution of better than 0.3 mm. For both CamE and CamP, the readout noise level corresponds to approximately a single analog-to-digital-converter (ADC) count.

## 2.2 Scintillator

The proprietary scintillator used in the FBSM device is generically referred to as Hybrid Material (HM). HM is an inorganic-polymer material available in both thin sheets and large area sizes. It is a fast scintillator with a decay time an order of magnitude less than the shortest exposure times of 50 $\mu$s, and hundreds of times less than the 1 - 10 ms proton beam dwell time in a single location, so the HM afterglow contribution to a dosimetric measurement is negligible.  In test bench experiments [16] (also see below) using a collimated beam of electrons from a $^{90}$Sr source, the HM scintillator produces a much larger amplitude signal per unit thickness than a smooth sur-faced but unpolished CsI(Tl) single crystal of the same area dimensions (see Section 3.1.1). It is noted that unpolished CsI(Tl) generates more front-surface emission than polished CsI(Tl). Im-portantly the beam image from the HM scintillator was clean without spurious reflections or vis-ible "blooming" around the edges (see Figure 4). The HM photon production efficiency is esti-mated to be similar to CsI(Tl), which is in the range of 48,000-65,000 photons/MeV [30, 31]. Being polycrystalline in nature, it is visually opaque and incapable of total internal reflection, thus re-sulting in a higher percentage of photons escaping from the front surface and with no appreciable back surface reflections and scattering. Unlike CsI(Tl), HM is non-hygroscopic and, as will be shown, exhibits excellent radiation hardness.

## 2.3 Fast Real-Time FPGA based DAQ

One of the important functions of the FBSM is to assert a beam interlock when the irradiation deviates from the dose delivery program. A relevant regulation, IEC 60601-2-1 [32], indicates that no more than 10% of the prescribed dose, or about 0.8 Gy (assuming the FAST-01 clinical trial as a benchmark [7] ) is delivered after interlock assertion. This demands both good dosimetric pre-cision and fast data processing. In order to meet the stringent IEC standard, the FBSM DAQ and analysis are implemented in Verilog firmware that runs on an FPGA evaluation board. A diagram of the hardware data flow is shown in Figure 2. Serial data from the cameras (i.e., types CamE or CamP) are processed on high bandwidth (12 Gb/s) lines via a dedicated interface, and multiplexer board to FPGA receivers that convert the serial bit stream to parallel data words. The firmware performs pre-processing operations such as background subtraction, and pixel amplitude correc-tions that remove non-uniformities. All processed image data are buffered locally. The data anal-ysis steps include beam finding, calculation of the centroid coordinates, beam 2D-widths and in-tegrated signal. The results of the analysis are ultimately compared to the clinical treatment plan specifying the planned dose, D(x,y), at each beam location. The clinical program will be mapped to the camera coordinate system and uploaded to the FPGA. Deviations of the measured and program dose would trigger a beam interlock (e.g., TTL) gate signal. A beta version of this firmware has been implemented and tested using a 1 $^{cm}$ diameter stepper motor position-controlled light source to emulate a beam. Results are reported below.

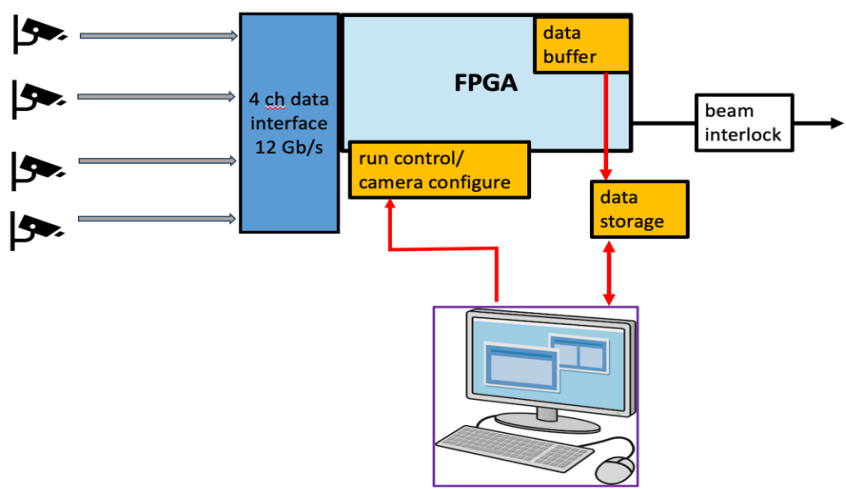

Figure 2: Schematic of data flow in the FPGA based DAQ and analysis system.

## 2.4 Calibration System

A three-part FBSM internal stability monitoring/calibration system is being developed.

Part 1 of this system corrects for the spatially dependent non-uniformities in the signal amplitude produced by a beam impinging on different regions of the scintillator. These non-uniformities are dominated by the geometric $1/r^2$ dependence on the optical path length from scintillator to the camera lens. Other contributions may come from fixed noise patterns in the sensor, pixel-to-pixel light sensitivity variations, and optical aberrations from vignetting and lens system astigmatism. The combined image non-uniformity is directly measured using a pre-calibrated flat-field luminescent screen, which mounts in place of the scintillator. Pre-calibration refers to removal of any intrinsic non-uniformities on the flat field source. These are mapped from large distance image captures insensitive to $1/r^2$ and cross-checked by acquiring images at $90^0$ rotations. Specifically, an image capture of this calibration screen is used to generate a pixel amplitude correction matrix normalized to a central reference point. This correction matrix has a mesh size of the order of 1 mm, selected to be much finer than the > 1 cm scale size of all expected optically induced variations. This matrix will be implemented in the FPGA firmware as a look-up-table and used to correct pixel ADCs to the reference point. The ultimate precision of this technique is the focus of ongoing work.

Part 2 of the calibration system uses an external, 50 KV, $41^0$ opening angle, unfiltered x-ray source using a tungsten anode (Moxtek, Inc.) on a fixed mount point to monitor the long-term scintillator response to radiation. This source, measured to have temporal stability of 0.1% RMS over 85 hours of continuous operation [33], will be used to generate a pixel field map for new scintillator screens. Images acquired after a preset minimum radiation exposure and/or after a fixed period, are used to determine the spatially differential degradation from clinical use, and to flag scintillator replacement should the degradation exceed a threshold. This application is insensitive to long term source stability. A detailed estimate of this fixed time period is reported in Section 4.4.

Part 3 converts the measured pixel signal to an absolute dose calibration by a method described in Section 4.5.

## 2.5 FBSM Prototype Performance Tests

We report here on the performance of a prototype FBSM having a maximum 15 cm × 15 cm sensitive area, and 12 cm longitudinal (along the beam axis) footprint. In these results a 10 cm × 10 cm scintillator area was used. These dimensions were dictated by two considerations: First, the prototype corresponds to a design that is small, light and easily transportable. Secondly, the large area coverage enables it to be potentially staged in future clinical trials.

The scintillator is viewed by the side-mounted camera through a front surface silver mirror (reflectivity >98% at 400-2000 nm) at an approximately $40^0$ angle. For the pulsed electron beam tests reported here a type CamE camera was used in beam triggered or asynchronous modes. Due to the viewing angle, the image is foreshortened along one coordinate axis and a homography transform or mapping algorithm is used to generate a final radiation dose map or display images in the beam coordinates. However, the FBSM validation of a clinical dose program in real-time does not require that any image collected during treatment be transformed. Such operations unnecessarily consume many FPGA clock cycles. Rather, it is the clinical treatment plan specified dose in beam/patient coordinates that gets transformed into the sensor coordinate system. This is a single operation, although not encoded in the firmware beta version described here, that can be easily done during DAQ initialization without any delay in analyzing the images.

A radiation shield enclosure is employed during FBSM staging in electron beams. This shielding is intended to reduce bremsstrahlung photons generated by electrons interacting with materials after emerging from the beam exit. These photons can directly hit the sensor of a CamE type camera and generate background. The GEANT4 [28] Monte Carlo simulation package was used to optimize the mixture of shielding materials that could fit in the limited envelope afforded by the prototype FBSM. The shield design consists of a two-layer box that surrounds the camera. The inner layer is 1 cm thick Pb, nested in an outer 3 cm thick shell of graphite. The low Z of the graphite impedes electrons while minimizing local generation of bremsstrahlung photons, while the high Z Pb is effective at photon absorption.

## 2.6 Laboratory Test Bench

### 2.6.1 Comparison to a reference scintillator

Evaluations of the HM scintillator were performed in a dark box mounted on an optical table and irradiated using a 3 mm collimated beam of $\beta^-$ particles from a 2.4 mCi $^{90}$Sr source. This source was positioned directly on the distal (back) face of the scintillator specimen. In these experiments, the FBSM camera type CamE recorded beam images on 2 cm × 4 cm, 0.43 mm thick HM scintillator, thicker than that used in the prototype FBSM, but otherwise structurally and chemically identical. A series of ten background images were collected without the source in place. The average of the backgrounds was subtracted from all signal images. Comparisons were made with respect to a reference 1.25 mm thick, smooth surfaced but unpolished CsI(Tl) scintillator with 1 s exposures, corresponding to roughly 0.002 Gy.

### 2.6.2 Spatial resolution for pulsed electron beams

Spatial resolution measurements were staged on an optical table. The FBSM was fitted with camera type CamE and was aligned relative an XY stepper-motor drive gantry upon which the above $^{90}$Sr source was mounted. The camera operated at a maximum resolution of 2 MP, compatible for operation to 100 Hz in pulsed electron beams. Resolutions at a higher pulse rate up to 1 kHz were obtained by re-binning the pixels into supercells of 2x2 and 4x4. The source collimator borehole exit was positioned within 1 mm of the surface of the scintillator and then stepped in increments of 1.000 mm ($\pm 1$ μm), horizontally across the width of the scintillator. At each position a 1 s exposure image was recorded for offline analysis of the beam centroids. The reconstructed centroids were plotted against the source position and fit by linear least squares regression. The resolution was determined from the RMS width of the fit residuals.

### 2.6.3 Fast readout tests

A beta version of the FBSM data acquisition and analysis has been implemented in firmware for real-time readout of the camera type CamP. The firmware performs all necessary operations including background subtraction, beam locating and calculation of the centroids, beam width and integrated signal. The total response time of the readout algorithm was measured with the Vivado FPGA logic analyser (Xilinx, Advanced Micro Devices, Inc., Santa Clara, California). Measurements of the total latency were done by injecting a trigger pulse to initiate the camera frame acquisition. The times to (1) acquire the frame, (2) receive all data in the FPGA and convert from serial to parallel data, (3) conduct all data processing such as background subtraction, and (4) perform analysis that includes calculation of centroids and RMS widths, were all recorded in units of FPGA internal clock cycles. A test of high data rate at 20,000 frames per second (fps) firmware analysis was done by robotically translating a 1 cm diameter LED emulated beam along one camera axis in 1-5 mm increments. At each position the camera collected and analysed approximately 30 events, and the analysis computed the beam centroids.

## 2.7 Prototype FBSM at UMH Radiation Oncology

The clinic facility includes a Varian TrueBeam™ electron linac. This machine provides 6-16 MeV electrons delivered in 2-4 μs pulses at time-averaged dose rates of 0.017-0.17 Gy/s. Beam control was done via the TrueBeam console using pre-programmed standard clinical settings. The specified dose was calibrated by UMH using the AAPM TG-51 protocol [34]. Absolute calibration was performed in water with source-surface-distance, SSD = 100 cm by ion chamber measurements at the depth of maximum dose: 6 MeV: 1.4 cm, 9 MeV: 2.1 cm, 12 MeV: 2.9 cm, 16 MeV: 3.9 cm with a 15 cm x 15 cm cut-out in the electron cone. The prototype FBSM was staged on a patient table, directly at the exit of the collimator positioned in the standard electron cone mounting structure as shown in Figure 3 (left). Data were acquired asynchronously with the beam pulses using 1 s long frames. A 5 mm diameter beam was produced using a tiered structured collimator shown in Figure 3 (right).

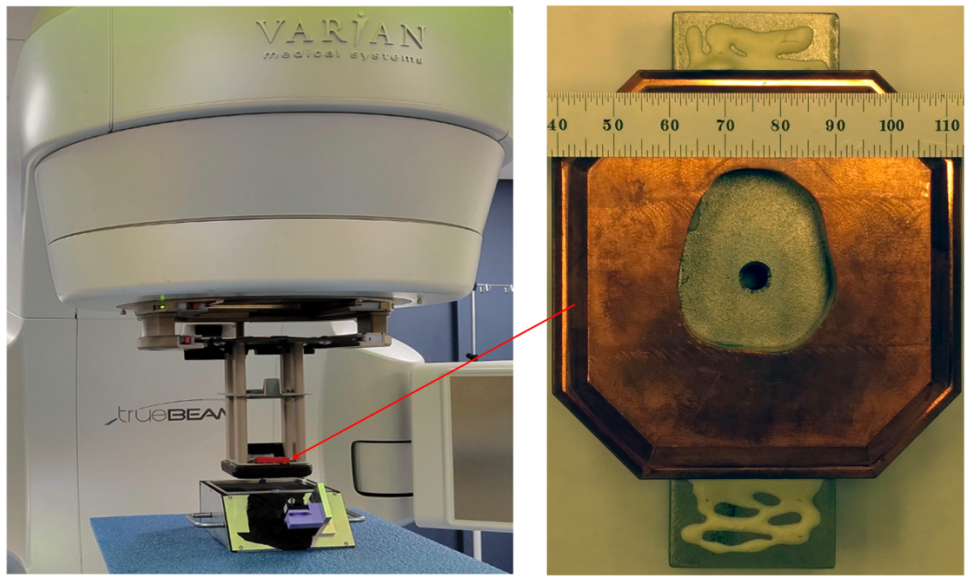

Figure 3: (left) Placement of the FBSM prototype device on the patient table, underneath the radiotherapy beam at UMH, Radiation Oncology. (right) Detail of the beam collimator structure used for the FBSM tests.

### 2.7.1 Imaging a radiotherapy beam

The initial experiment done at UMH was intended to demonstrate beam imaging capabilities in a clinically relevant setting, and to evaluate backgrounds from bremsstrahlung radiation. Images were collected at 6 MeV and 16 MeV, and at 1 Gy/min to 10 Gy/min dose rates. However, the background studies used the 10 Gy/min dose rates.  Images were also compared to that acquired by Gafchromic film.

The effects of backgrounds were assessed by measuring the distribution of ADC counts in remote regions of the sensor field that do not directly view the scintillator during irradiation.  Pixel hits in these regions can be caused by photoelectric interactions of x-ray photons in the camera sensor.

### 2.7.2 Gafchromic film

Gafchromic films are a standard instrument used in RT clinics for dose measurement and quality control.  Notably Gafchromic[TM] EBT-XD film (Ashland Inc.) has a maximum dose limit of 40 Gy. It produces sharp images and a manufacturer specified intrinsic spatial resolution of 25 µm. The Gafchromic film images used were 10 cm × 10 cm and were digitized with an Epson 10000XL RGB color flatbed scanner at 72 dpi and 200 dpi. The Gafchromic data were prepared and analysed by UMH Radiation Oncology using commercial proprietary software (FilmQA Pro 2016 v5.0, Ashland

Inc.) and using a calibration procedure that corrects for a non-linear optical density to dose response, and film and scanner non-uniformities using the three scanned color channels [35].

In the experiments conducted on the Varian TrueBeam, a 100 cm$^2$ square film sample was positioned on top of the FBSM scintillator bed, such that its vertical distance to the beam exit (about 40 cm above) was the same as the scintillator to about 0.1 %. The film was irradiated with a 16 MeV electron beam set to a 10 Gy/min dose rate and exposed for 2 minutes, such that the film received a 20 Gy equivalent dose delivered to the hypothetical calibration point.

### 2.7.3 Bremsstrahlung backgrounds

The position of the FBSM near the radiation field made it potentially vulnerable to direct bremsstrahlung x-ray background hits on the pixels. These backgrounds are nominally produced by interactions of the electrons off the tungsten "jaws" in the TrueBeam machine, or in downstream materials. The jaws shape the broader electron beam field before final shaping by the electron cone and collimator for tumor irradiation. Evaluation of these backgrounds was done measuring the average ADC in a selected region of 200 x 200 pixels in a "dark" corner of the sensor field. This dark corner was not exposed to the scintillator or any source of light, and therefore signals above the normal readout background noise were created only by direct hits of background photons on the pixels. This background was measured using 6 MeV and 16 MeV electrons at 10 Gy/min dose rates with and without the graphite/Pb radiation shield surrounding the camera. These conditions of highest energy and dose rate represent the most severe that could be generated using the Varian machine.

## 2.8 FLASH-Compatible Dose Rates in the NDRL Electron Beam

The NDRL beam consists of 8 MeV electrons arriving in short, $\delta t_{pulse}$ = 1 ns wide (FWHM) pulses at a repetition frequency, $f_{rep}$ = 30 Hz. A beam synchronization signal was used to trigger the DAQ. The leading edge of this trigger pulse arrived about 3 $\mu$s before the 1 ns beam pulse. The camera shutter time was set to 1 ms, leaving most of the exposure window dark. Before each run a series of beam-off dark frames were collected, averaged, and subtracted from the signal frames. The prototype FBSM was positioned at approximately 5 cm from the beam line exit window. Beam collimation was achieved with a 0.5 cm bore in a 2.5 cm thick steel plate mounted in front of the beam exit window. The total beam current was initially set using an inline inductive pickup located about 3 m upstream before the focusing quadrupole magnets. The beam current exiting the beam pipe incident on the scintillator was measured with a Faraday cup (FC) electrode. This electrode consisted of a 6.3 cm (2.5 inch) diameter steel disc that was mounted on a remote-controlled stepper motor translation arm. The FC was connected to a charge integrator/amplifier with a 4.3 nC/V voltage-to-charge transfer function and read out on a digital sampling oscilloscope (DSO). The dose per pulse (DPP) was determined from the product of the beam intensity and energy loss in water (or in tissue equivalent material): $DPP = \phi \frac{dE}{dx}$ where $\phi$ is charge fluence in units of nC/cm$^2$-s and $\frac{dE}{dx}$ is the electron stopping power in units of MeV cm$^2$/kg [29]. The difference in energy loss of tissue equivalent material, water and the proprietary HM is less than 3%. The effective beam diameter was defined as the full width at ¾ maximum, (FW3QM) of the beam spot. This limited region near the peak was selected because it receives the highest dose. The beam spot and profile are shown in Figure 11 left and right panels.

Two measurements were performed at NDRL: (1) radiation hardness of HM scintillator and (2) signal response vs dose.

### 2.8.1 Radiation hardness

The radiation hardness measurement was done by setting a high DPP = $8.0 \pm 0.4$ Gy/pulse, $f_{rep}$ = 30 Hz and acquiring data continuously for ¼ hour. The total time averaged dose rate <dD/dt> = $240 \pm 12$ Gy/s and the total cumulative dose, $D_{tot}$ = 216 kGy. The source of the uncertainty is described in the Results section. The HM light yield was measured in the central region of the beam by monitoring the pixel charge in ADC counts. This signal was expressed relative to a control scintillator region away from the primary beam that received a very low dose of < 1 % of the primary beam region. This signal region/control ADC ratio removes most of the longer-term variations in the beam current.

### 2.8.2 HM response to dose

The dose response of the FBSM instrumented with HM scintillator was measured by increasing the charge per pulse of the NDRL beam, from about 0.11, 0.22, 0.43, 0.92, 1.36, 1.70, 3.2 nC. Immediately following each change of the pulse charge, the FC was momentarily positioned in the beam path to measure the time averaged current, and then retracted so that the beam impinged on the FBSM entrance window. The uncertainties associated with this method are discussed in the Results section.

## 2.9 Heavy Ions at FRIB

Experimental results were collected at FRIB using a scintillator-based ion beam monitor (SBM) designed for real-time operation in high vacuum beamlines. This SBM, incorporating similar optics, camera and scintillator materials also served as the initial FBSM development platform. These tests included a sample of 205 μm thick HM scintillator and camera type CamE. The FRIB beam consisted of $^{86}$Kr$^{+26}$ ions at 2.75 MeV/u running at currents from a few particles per second (pps) to $5 \times 10^5$ pps, and which 100% of the ion energy (236 MeV) was dissipated in the scintillator. The dose rates acquired at FRIB for the highest beam currents extended to ~50 Gy/s, rendering the results relevant to FLASH applications. Three important FRIB results previously reported in conference proceedings [16] are:

(1) The SBM using HM scintillator is sensitive to single ions.
(2) Linear response: The beam current was increased from 5 pps to 520,000 pps. The beam current was measured by integrating the total light signal in all pixels comprising the beam spot and normalized to the average of integrated signals of all isolated ion hits collected in the lowest current data sets. Over a range extending five orders of magnitude, the signal response of the SBM to beam current was linear to within the uncertainties available by the independent current measurement instruments.

(3) Real-time operation: The ion beam monitor was also configured with fast, online software that ran on the computer's DAQ to provide analysis and display in real-time, at ~ 1 Hz update rates. The same basic beam analysis algorithms have now been incorporated into much faster FPGA firmware for a FLASH compatible beam monitor.

# 3  RESULTS

In the following section we report the results of the above-described experiments, grouped by facility for clarity.

## 3.1 Laboratory Testbench

The corresponding methods for these results are described in Section 2.6.

### 3.1.1 Comparison of HM to CsI(Tl)

Previously reported preliminary work established that HM produced a very clean image, free of reflections and edge smearing effects from photon blooming or scattering and with significantly larger amplitude signals than observed in the benchmark CsI(Tl) crystal [16].   New results reported here validate this observation. Figure 4 (left and right panels) respectively show the background-subtracted image generated by the source in the CsI(Tl) crystal and for the HM scintillator sample. The average signal amplitude is represented by the mean of the ADC spectrum (shown on the Z axis log scale) in the regions defined by the square boxes in the figures. The result is 97±0.14 and 237±0.22 ADC counts for CsI(Tl) and HM respectively. The two scintillators have approximately the same density, so when normalized to unit scintillator thickness in mm of material or mm WET, the HM produces about a 7.1 times larger signal for this test source. This is also shown in the beam profile projection plots in Figure 4 (right). Here the average ADC of a 40 pixel wide band is plotted along the center of the beam spot for each scintillator type. The ADC count is then normalized to the scintillator thickness to show the relative signal yields. A log Y scale is used to show the off beam tails, and spurious reflection effects in the CsI that are not present in the HM.

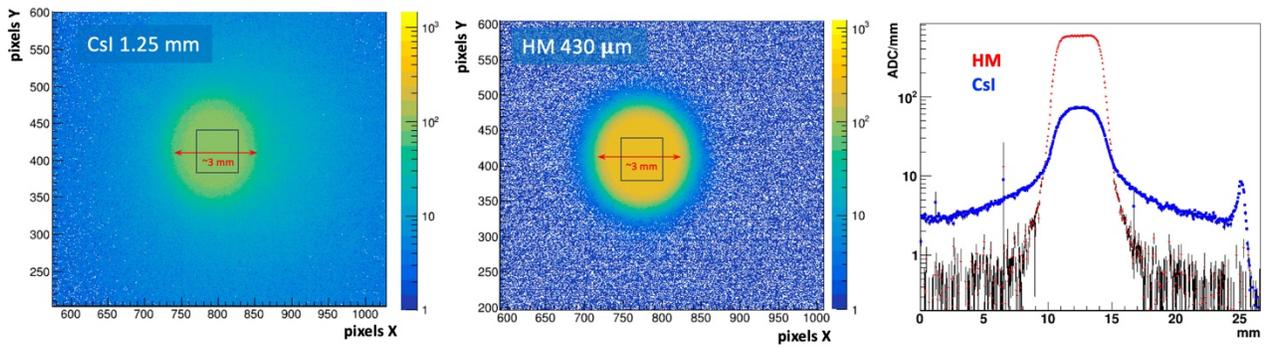

Figure 4: Background subtracted images of a 3 mm collimated β⁻ particles on 1.25 mm thick CsI(Tl) crystal (left) and 0.43 mm thick HM scintillator (center). Projection histograms (right) shows the beam profile averaged over a 40 pixel wide band along the orthogonal direction. The Y axis is the ADC count normalized to the scintillator thickness.

The strong signal yield of HM scintillator is attributed to its fine granular nature, and absence of total internal reflection that is common to single crystal scintillators or bulk plastic scintillators. This fine granularity of the scintillation domains also produces a visually "clean" image with well-defined boundaries. This is demonstrated in Figure 4 (center and right) showing a beam spot (measured at the full width at 10% maximum) that corresponds to the size of the collimator diameter used, and with background levels nearly three orders of magnitude below the signal.

### 3.1.2 Spatial resolution

Figure 5 (left) shows the reconstructed centroids plotted against the precise location of the source, determined by the stepper motor position, with a linear fit superimposed. The residual distribution of the reconstructed positions relative to the precision source position is shown in the right panel.

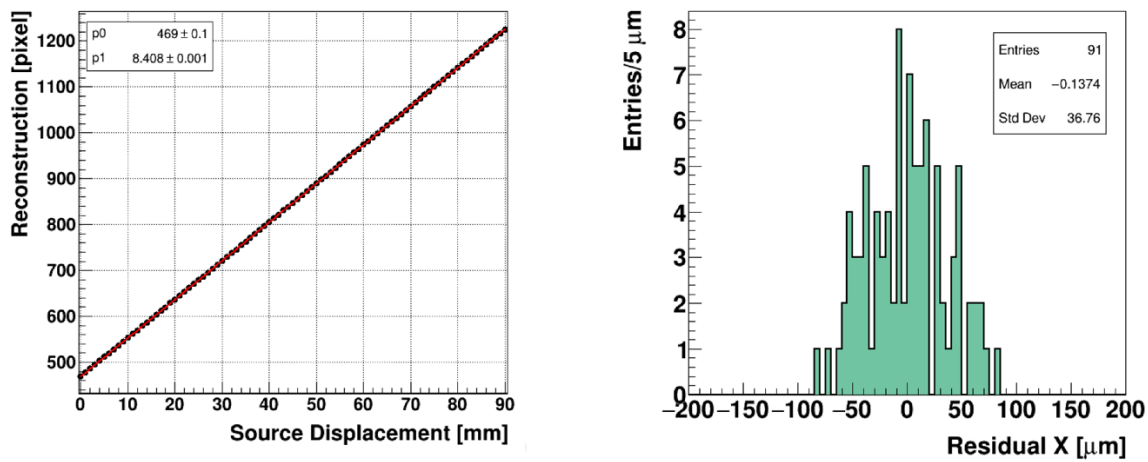

Figure 5: (left) reconstructed beam position in pixel units of a 3 mm beta source translated along the X coordinate of the FBSM. The data points are in black, and the red line is a linear fit. (right) The residual distribution of the reconstructed positions.

The spatial resolution, represented by the RMS width of the fit residuals, equals 37 μm. For comparison, the intrinsic manufacturer specification of the spatial resolution of Gafchromic film is 25 μm [36]. The spatial resolution measurement reported here corresponds to operating conditions for a pulsed electron beam, using a camera type CamE at a maximum frame rate of 200 fps, in which the full set of two million pixels are read out. A higher frame rate up to 1000 fps can be achieved by the use of an updated camera type CamE, combined with coarser pixel binning, read out in 2x2 or higher groupings. The corresponding spatial resolution at the higher frame rate was estimated by re-binning the pixel data into 2x2 cells enabling a frame rate of 1000 Hz. For reference the resolutions of coarser groupings of 4x4 and 8x8 cells, were investigated. The results of this regrouping exercise are listed in Table 1.

Table 1

| Estimated resolution vs frame rate for a ~3 mm beam. | | |
|---|---|---|
| Binning | Max frame rate (fps) | Resolution (μm) |
| 1 x 1 | ~200 | 36 |
| 2 x 2 | 1000 | 38 |
| 4 x 4 | 1000 | 39 |
| 8 x 8 | 1000 | 44 |

The modest sensitivity of the resolution to the pixel binning is attributed to the size of the beam spot, including its tails, which is projected over at least 1000 pixels, allowing ample charge sharing and precise centroid determination. Beyond 2x2 binning there is no increase in frame rate.

3.1.3 FPGA DAQ results

As noted, the total response time of the readout algorithm includes the frame acquisition, data transfer, data processing and analysis. Figure 6 was generated by a virtual logic analyzer that was instantiated inside the FPGA firmware. This timing diagram is from the image processing of camera type CamP. It includes the sensor data transfer time and analysis times as measured by the virtual logic analyser. Refer to caption for details. The analysis firmware pre-processes the data during the transfer and starts the beam finding, centroid and width calculations immediately after the end of the frame acquisition. XDIV_DONE marks the total time need by the firmware to analyse the data: for reference, the rightmost vertical line in the central inset shows a delay of 1 μs from the end of the frame acquisition.

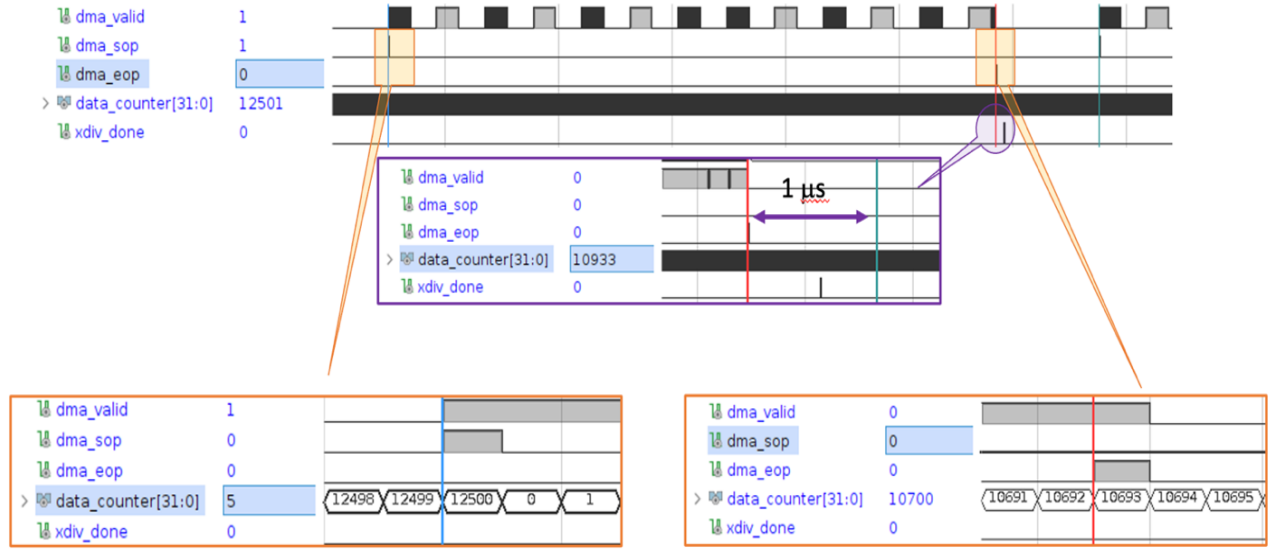

Figure 6: Firmware beta version timing diagram from FPGA logic analyzer. The top timing chart shows a sequence of valid data blocks (DMA_VALID) comprising a single camera image frame. They are bookended by a start of image frame (DMA_SOP) and end of image frame (DMA_EOP). DATA_COUNTER registers the number of internal 4 ns FPGA clock cycles. The time between two sequential DMA_SOP markers is 12500 clock cycles corresponding to 50 μs. The bottom left inset shows the detailed timing of the DMA_SOP signal, which is used to reset DATA_COUNTER. The bottom right inset shows the detailed timing of the DMA_EOP which occurs approximately after 42.7 μs of the total of 50 μs between frames. The middle inset shows the details of the XDIV_DONE signal, marking the end of the frame analysis relative to the DMA_SOP, and that the analysis is completed in about 0.7 μs.

Table 2 summarizes the data acquisition processing time results with the camera running at 20,000 fps. The times shown are incremental. The frame data reaches the firmware after a 50 μs exposure and is transferred to the firmware while the camera collects the next frame. The real-time FPGA firmware performs pedestal subtraction, data correction and pixel ADC weighted sums during the transfer time. After the frame has been fully received, initial beam position centroids are calculated. The final beam position and width are computed using a reduced pixel box (17 x 17 for this test) around the beam centroid. The total frame processing time is the firmware latency from the end of the frame capture to the conclusion of the final beam variable analysis.

Table 2

| DAQ step | Frame duration | Frame transfer time | Beam finding | XY beam position | XY RMS widths | Total frame processing time |
|----------|------|------|------|------|------|------|
| Time | 50 μs | 42.8 μs | 272 ns | 304 ns | 188 ns | 43.6 μs |

Figure 7 (left) shows the firmware generated LED beam centroids as a function of 33 beam positions, and Figure 7 (right) shows the residuals. This result indicates that operation at 20,000 fps using camera CamP achieves a resolution of 67 μm for a 1 cm beam.

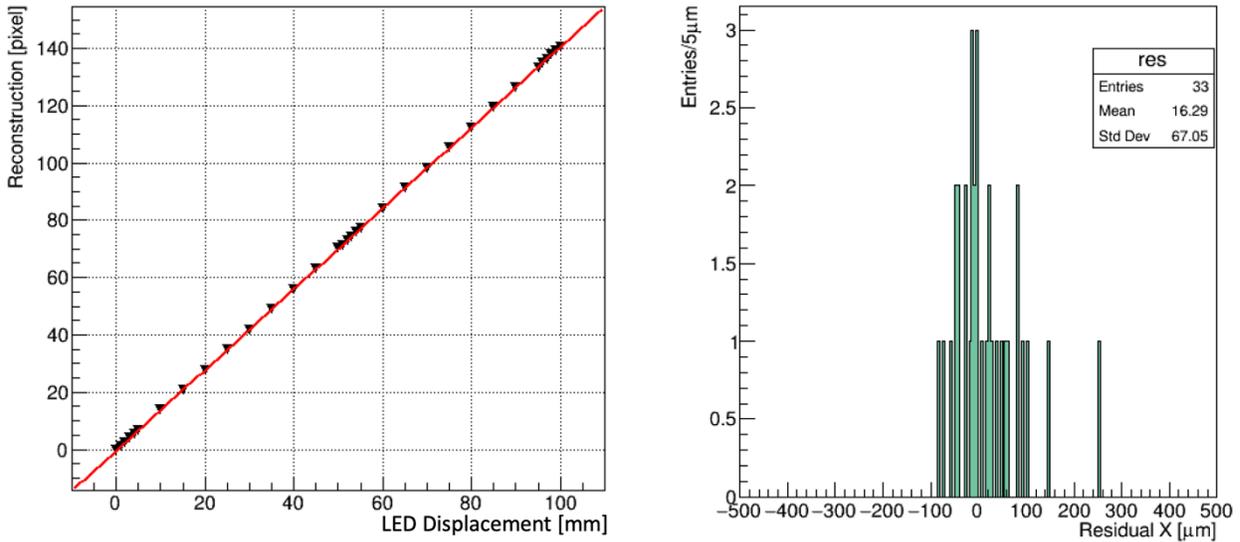

Figure 7: (left) The centroids, reconstructed in firmware, of a 1 cm LED emulated "beam", translated along the X coordinate of the FBSM operated at 20,000 fps. Data points are in black; the red line is a linear fit. (right) The residual distribution of the reconstructed positions.

## 3.2 Prototype FBSM at UMH Radiation Oncology

The corresponding methods for these results are described in Section 2.7.

### 3.2.1 Beam visualization

Figure 8 (left) shows the 1s exposure background-subtracted image in camera coordinates for a 16 MeV beam delivering 0.17 Gy/s. The trapezoidal shape and rectangular aspect ratio derives from the oblique angle view of the camera on the scintillator. Figure 8 (center) shows this same image after a homography transform processed using OpenCV. The false-color palette reveals the primary electron beam and also the structure of the non-uniform mass profile of the collimator shown in Figure 3 (right). In particular the inner tumor-shaped aperture and the octagonal perimeter of the Cu collimator are readily apparent, with ADC levels at < 10% of the primary beam. The average signal in the peak region of the beam is 2800±100 ADC counts, corresponding to $1080 \pm 40$ photoelectrons (PE).

## 3.2.2 Comparison to Gafchromic film

The right panel of Figure 8 shows an image produced by this same beam for a 2 minute exposure to Gafchromic EBT-XD film, corresponding to a 20 Gy dose. The film image hints at the overall collimator structure, albeit with much less clarity than the FBSM. Importantly the FBSM image is collected and can be processed in real time whereas the film processing is done over an hours-long time scale.

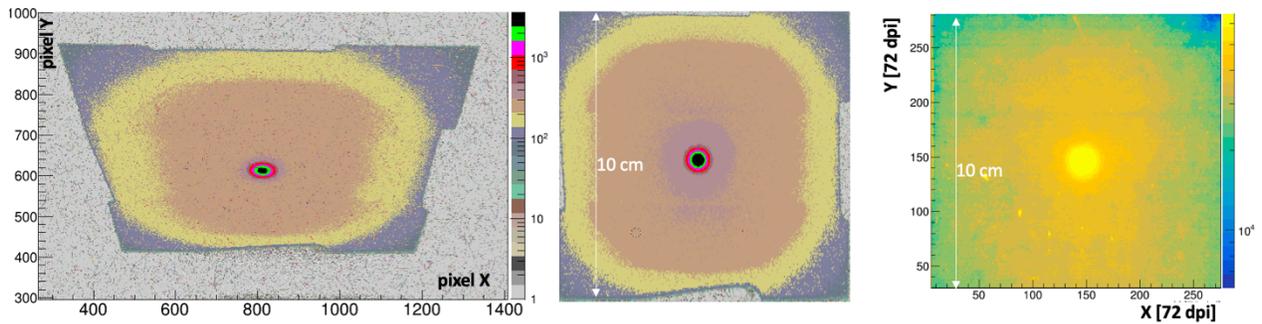

Figure 8: 16 MeV electron beam: (left) FBSM image, exposure = 1s (0.17 Gy), in sensor pixel coordinate system, original aspect ratio; (center) scintillator image after a homography transform; (right) Gafchromic film image for 20 Gy isocenter-equivalent dose exposure. Note that the 10 cm scale refers to total active area dimension. The 7cm dimension of the Cu collimater shown in Figure 3 is outlined by the inner, darker tan region.

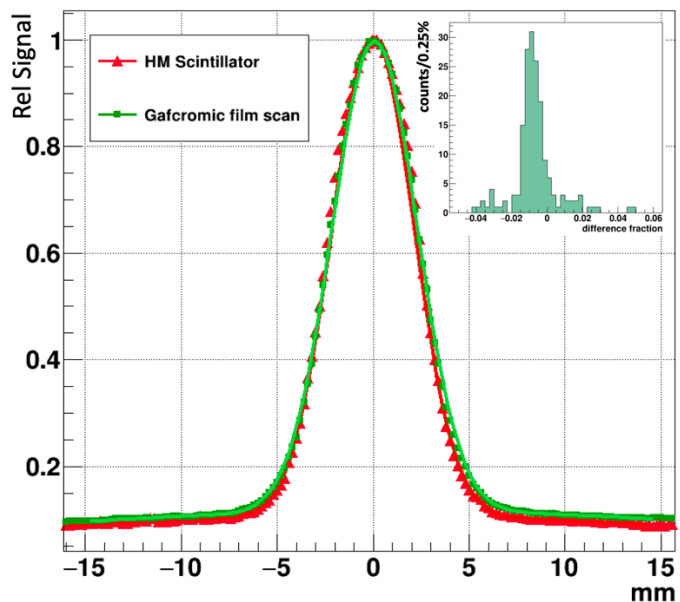

Figure 9: The projection of the beam along the X-axis for an averaged 10 pixel wide Y-axis band for the Gafchromic film and prototype FBSM image data. The double Gaussian fits of each curve are shown. Inset: residual difference histogram of the two curves. The RMS width is 0.5%.

The relative average signal amplitude for a 10 pixel-wide band projected along the X coordinate and passing through the beam center is shown in Figure 9 for the Gafchromic film overlayed with the prototype FBSM image. The beam profile plots were each fit with double Gaussian distributions of the form: $f(x) = A_n e^{\left(\frac{x}{2\sigma_n}\right)^2} + A_w e^{\left(\frac{x}{2\sigma_w}\right)^2}$, where the subscripts $n$ and $w$ denote the narrow primary beam and the wide tails, respectively. The fit results for HM and Gafchromic (in parenthesis) distributions are: $A_n$=0.92 (0.9), $\sigma_n$ = 2.11 (2.18) mm, $A_w$ =0.106 (0.113), $\sigma_w$ = 26.2 (30), where the fit errors are represented by the last decimal place. That is, the two distributions can be described by the same functional form where the parameters fall within 1.1%-6% of the averages of the two fits. The residual or difference distribution, shown in the inset, has an RMS spread of 0.5%. These results indicate the fidelity of beam imaging of the prototype FBSM is comparable to Gafchromic film.

### 3.2.3 Backgrounds from bremsstrahlung

The background hit occupancy is visualized in Figure 10. Each panel shows the same 100 pixel x 100 pixel field in a dark corner of the images, unexposed to any light emanating from the scintillator. Panels (A) and (B) respectively show an exposure with 6 MeV and 16 MeV electrons without any radiation shield, while panels (C) and (D) are with the shielding installed. While a qualitative reduction in background hit occupancy provided by the shield is clear, the net efficacy on background reduction is, for this work, best represented by the ratio of background to signal ADC count, or equivalently the pixel charge (averaged over all pixels in the dark corner regions and over the beam regions), reported in Table 3.

Table 3

| Shield present | Energy [MeV] | Bkg [ADC] | Signal [ADC] | Bkg/Sig | Shield reduction factor |
|---|---|---|---|---|---|
| Effect of shielding on 6 MeV and 16 MeV electron beams at 0.17 Gy/s. ADCs are averaged 1 s exposures over the beam core for the signal, and over the four "dark" corners of the pixel sensor for the backgrounds. | | | | | |
| No | 6 | 27 | 629 | 4.3% | -- |
| Yes | 6 | 6 | 642 | 0.9% | 4.8 |
| No | 16 | 54 | 2299 | 2.3% | -- |
| Yes | 16 | 11 | 2411 | 0.4% | 5.8 |

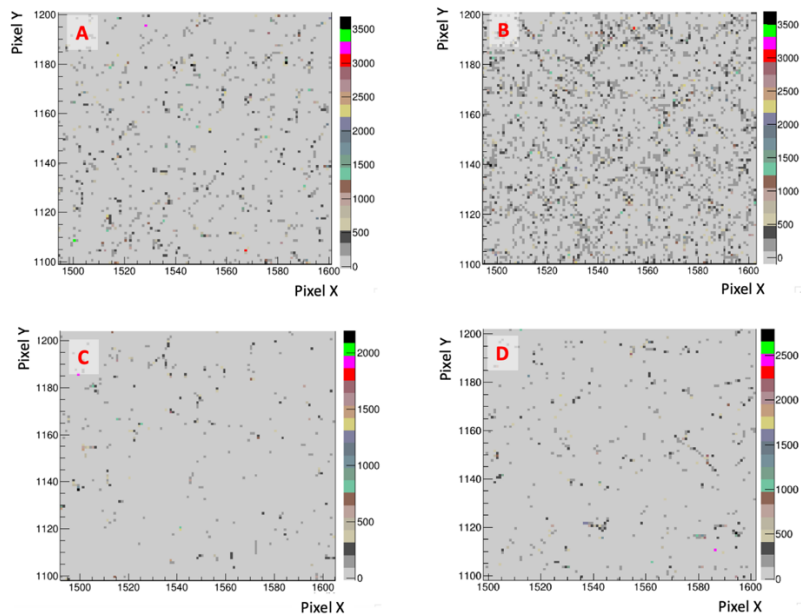

Figure 10: Bremsstrahlung photon backgrounds for (A & B): 6 MeV & 16 MeV electrons, no shielding; (C & D) Same, but with the shielding enclosure surrounding the camera.

The shielding reduces the background contribution, on average, by a factor of 5.3. Importantly, with the shielding in place the average contribution of this background to the signal is less than 1%. This background, when averaged over many thousands of pixels, effectively behaves like a constant pedestal offset and can be subtracted from signal data.

## 3.3 FLASH-Compatible Dose Rates in the NDRL Electron Beam

The corresponding methods for these results are described in Section 2.8.

Figure 11 (left) shows an image of a single 1 ns pulse, charge Q = 3.3 ± 0.17 nC, hitting the scintillator. Figure 11 (right) shows the beam profile along a horizontal axis, and the Full Width at ¾ Maximum (FW3QM) used for dose calculations in the following section. The NDRL beam current has three sources of statistical and systematic measurement uncertainties: (1) The pulse-to-pulse periodic fluctuations measured at the highest pulse setting used, and recorded in pixel ADC count, varied by ± 5 to an average value of 320, or 1.6%; (2) a 1% periodic variation in signal amplitude was observed over a time scale of minutes; (3) the FC charge was nominally measured by a pre-set DSO function that determined the voltage difference of a waveform plateau relative to a fixed reference baseline. This waveform baseline was observed to drift by approximately 5% based on the data. These uncertainties are independent and add quadratically. The combined uncertainty on the pulse charge/beam current measurements is 5.3%.

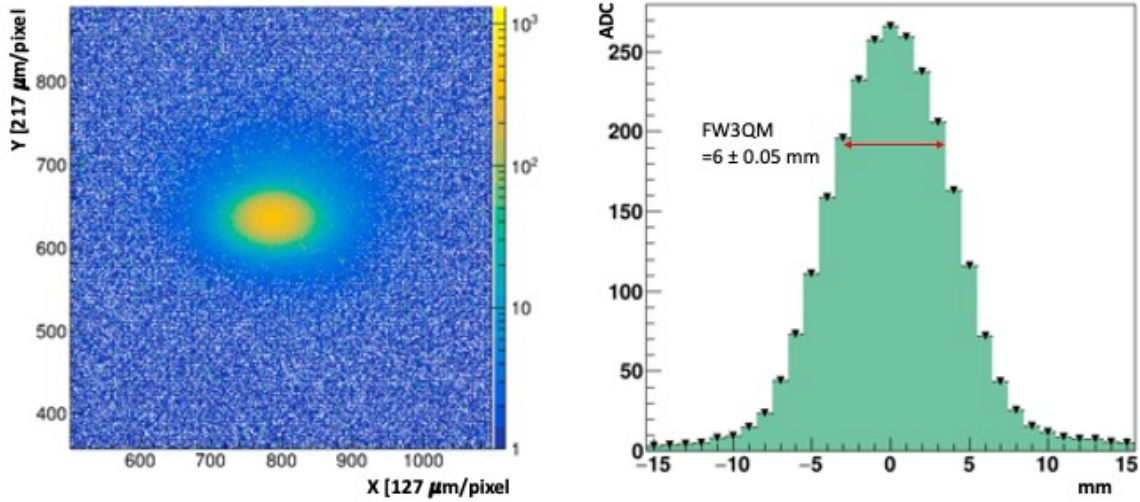

Figure 11: (left) image of a single pulse, 3.3 nC.  (right) Beam profile along X axis of the primary beam region.

### 3.3.1 Radiation hardness of HM scintillator

Using the same beam shown in Figure 11 above, Figure 12 shows the ratio $R = \frac{A(t)_s / A(0)_s}{A(t)_{con} / A(0)_{con}}$ where $A_s$, $A_{con}$ are respectively the average ADC signals in the primary beam signal and control regions at time t from the start of irradiation. (This method is explained in Section 2.8.1).  The control region is completely outside of the primary beam and receives about 0.3% of the beam current per unit area. For the signal region, as noted earlier, we consider the beam area contained by the FW3QM. The projection along one axis in Figure 11 (right) has FW3QM = 6 ± 0.05 mm. The normalized area of the FW3QM is 37%, the area of this region is 0.304 cm$^2$ and the charge per unit area: = 4.0 ± 0.2 nC/cm$^2$. The "ratio of ratios", R, removes some of the time dependent beam periodic fluctuations. The signal loss, obtained from a linear fit over the 216 kGy cumulative dose, is -0.025% (± 0.0015%)/kGy.

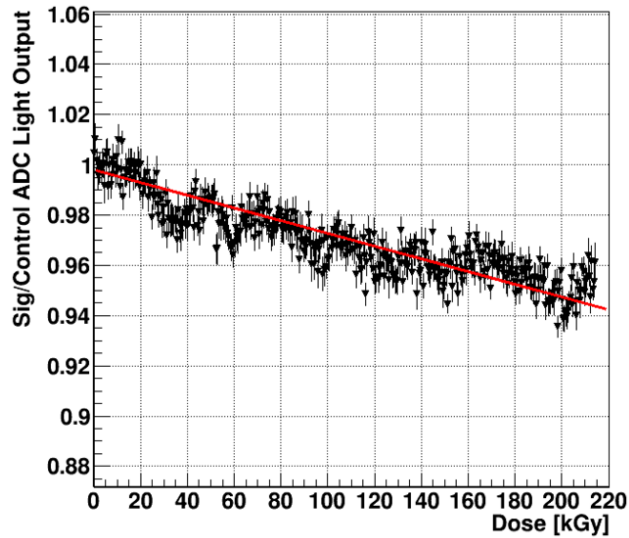

Figure 12: HM scintillator normalized signal during extended irradiation. The average slope indicates an average signal loss of 0.025 % per kGy exposure.

### 3.3.2  HM linearity

The average beam signal measured over 0.4-3.1 nC is plotted in Figure 13. The X coordinate is expressed in units of the equivalent DPP range of 0.28-7.9 Gy. We can express this dose rate range also as a time average dose rate <dD/dt> = DPP × f$_{rep}$ range of 8.4-237 Gy/s, thus bracketing a FLASH compatible regime. We also note that the *instantaneous* dose rate, D$_{ins}$ = DPP/δt$_{pulse}$ spanned from about 0.3 - 8×10$^9$ Gy/s due to the 1 ns FWHM pulse width. The red line fit indicates a linear response over the FLASH compatible region up to 4 Gy/pulse or 120 Gy/s. The highest data point at 237 Gy/s falls 6% below the fit line, exceeding, slightly, a single standard deviation indicated by the error bar. This anomaly will be investigated with a new set of more precise measurements that are planned.

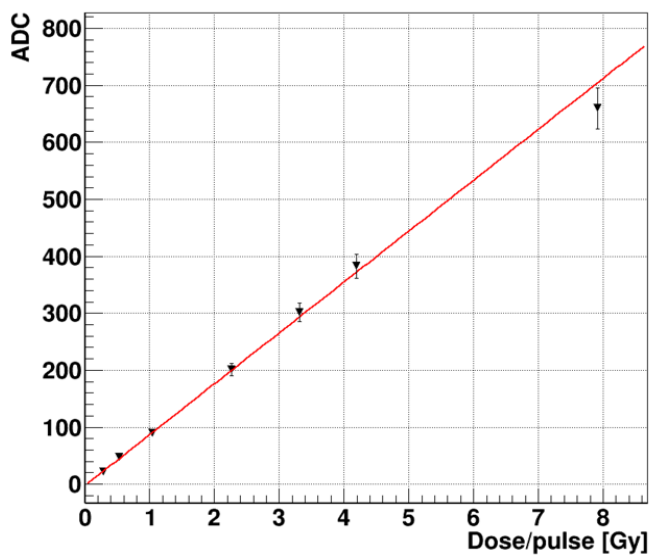

Figure 13: Prototype FBSM average ADC signal of 8 MeV electron beam vs radiation equivalent dose per pulse.

Similarly, results from FRIB [16] ($^{86}$Kr$^{26+}$ ions ΔE = 236 MeV total energy) show that the HM scintillator is linear to beam current for five orders of magnitude from ~5 to 520,000 pps, distributed over an irregular beam spot with an approximate area of 0.08 ± 0.01 cm$^2$ so that the maximum intensity $\phi$ = 6.5 ± 0.8 x10$^6$ pps/cm$^2$. SRIM-based calculations [37] of the ion range in HM scintillator material is Δx = 48 μm. The average dose deposited in this layer, density ρ=4.5 g/cm$^3$, is

$\phi$[pps/cm$^2$] $\times$ $\Delta$E/$\rho\Delta$x [ev/cm] $\times$ e [J/eV]$\times$1000 [g/kg] = 51 $\pm$ 6 Gy/s. Therefore, FRIB heavy ion data are also consistent with a linear HM scintillator signal response up to a FLASH compatible dose rate.

# 4  DISCUSSION

The experiments conducted with the FBSM prototype reported in this paper demonstrate attributes necessary for FLASH RT applications. In the following, we discuss the implications and limitations of these measurements, outline how the FBSM can be used in a FLASH-capable beam and comply with IEC requirements, and consider related beam monitoring technologies.

## 4.1 Fast FPGA Readout and its Application to FLASH

One of the important functions of the FBSM is to assert a beam interlock when the irradiation deviates from the dose delivery program. A relevant regulation, IEC 60601-2-1 [32] indicates that no more than 10% of the prescribed dose is delivered after interlock assertion. This demands both dosimetric precision and sufficiently rapid data processing of the FBSM. A beta version of the FBSM DAQ analysis, described earlier, has been implemented in firmware for real-time readout of the camera. The total response time of the readout algorithm was measured to be about 43 $\mu$s, although there is a net 93 $\mu$s lag of the analysis output from the start of the data frame due to a 50 $\mu$s delay in the initial frame's data processing. The IEC 60601-2-1 regulation nominally applies to pulsed electron beam delivery with as few as 10 pulses per treatment, where the pulse rate is hundreds of Hz, and always under 1 KHz. Since the FPGA firmware delivers an analysis in about 50 $\mu$s, we consider the more challenging case of a quasi-continuous proton beam from an isochronous cyclotron that delivers at a dose rate R. The residual dose allowed by the FBSM therefore, would not exceed R$\times$dT$_i$, where dT$_i$ is the ~50 $\mu$s time to assert a beam interlock. Assuming a nominal FLASH dose rate of R=100 Gy/s, the residual dose is 0.005 Gy. Consequently, a dose as small as 0.05 Gy could be prescribed at a 100 Gy/s dose rate, and therefore, the FBSM would be in regulatory compliance with such a FLASH application.

## 4.2 Backgrounds: From Conventional to FLASH RT

Results obtained from the FBSM operation at UMH reveal the performance features for pulsed electron beams in a clinical setting, albeit at conventional and not FLASH dose rates. The dose rates from NDRL, while compatible with electron FLASH, were delivered with a research beam and not a clinical RT linac, and so it is difficult to understand the true backgrounds. The primary implications are that in FLASH conditions both the signals and backgrounds from bremsstrahlung may be many times larger. Here, we extrapolated expected signals and backgrounds for various FLASH conditions.

The dose rates acquired at UMH at 100-1000 cGy/minute yielded an equivalent 0.017-0.17 Gy per untriggered image frame. The signal at 0.17 Gy/frame was 1080 PEs. In a FLASH delivery mode, a therapeutic dose is then delivered in a sequence of μs scale pulses that trigger the acquisition of a single image frame. We, therefore, considered the expected signal and background levels per frame for various FLASH delivery parameters. Furthermore, the GEANT4 optimized radiation shield keeps the background produced by conventional electron RT to < 0.9% of the signal for clinically relevant tests. Importantly, this background should scale with the dose per pulse, or per image frame. Table 4 shows the expected signal levels and associated bremsstrahlung backgrounds per frame for various FLASH delivery pulse rates, using as a benchmark 10 Gy total dose delivered at a rate of 40 Gy/s for 0.2 s. Similar results are expected using as a dose benchmark the currently recruiting IMPulse clinical trial [38], which delivers 30 Gy hypofractionated in three 10 Gy doses per fraction [39]. The final column is the ratio of this background to the nominal readout noise level of 2.3 e.

Table 4: Signals and projected backgrounds from bremsstrahlung photons on the camera sensor for a 10 Gy dose.

| Pulse rate [Hz] | # pulses | DPP [Gy] | Signal [PE] | Avg Bkg [PE] | Bkg/Dark |
|-----------------|----------|----------|-------------|--------------|----------|
| 100 | 20 | 0.5 | 3176 | 29 | 12.6 |
| 200 | 40 | 0.25 | 1587 | 14 | 6.1 |
| 500 | 100 | 0.1 | 635 | 6.2 | 2.7 |
| 800 | 160 | 0.0625 | 397 | 3.5 | 1.5 |
| 1000 | 200 | 0.05 | 318 | 3.1 | 1.3 |

We note that overall backgrounds decrease inversely with pulse rate since the same dose is distributed over more pulses. For 10 Gy, the added average background exceeds the readout noise, but in no case is it expected to rise above 0.9% of the signal.

While the bremsstrahlung backgrounds from electron FLASH appear understood and manageable, the background neutron hit rate produced by proton FLASH remains to be established. Investigations of neutron fluence in the proximity of a proton beam in a patient or phantom suggest low rates. Monte Carlo simulations that quantify neutron production by 100-250 MeV protons interacting in a water phantom --representing a patient-- as a function of the proton kinetic energy indicate neutron differential fluences for energy $\leq$ 10 MeV that reach a maximum for 250 MeV are of order $d\varphi/dE \sim 3\times10^{-6}$ MeV$^{-1}$ cm$^{-2}$ and fall off rapidly above 5 MeV [40]. For a conservative estimate we ignore this falloff and integrate to obtain $\varphi \sim 3\times10^{-5}$ cm$^{-2}$, noting that this is the fluence integrated over all scattering angles which peak in the forward direction, and not at the camera where the rates will be lower. We can estimate the number of fast, non-thermal neutrons incident on the camera sensor of area $A_{sens}$, for FLASH level proton beam current $I_p$ simply as $N_n = \varphi \times A_{sens} \times I_p \times \delta t$. For camera CamP, $A_{sens} = 0.07$ cm$^2$. The maximum average current is in the machine specific range $I_p = 30 - 800$ nA [41], and the frame time is $\delta t = 50$ μs, yielding a maximum of 20-524 neutrons per frame. The fraction, Q, that interacts to produce a background hit can be further estimated from the ratio $Q = \tau/\lambda$, where for camera CamP, $\tau = 0.075$ cm is the sensor thickness, and the interaction length $\lambda = 1/\sigma\eta$, $\sigma$ is the unweighted average neutron-silicon total interaction cross section [42] [43] at 1.5-10 MeV of 1.2 barns, and $\eta$ is the silicon number density = $5\times10^{22}$ cm$^{-3}$. This yields Q = 0.4%, so that the maximum background is in the range of 0.1-2.3 hits/frame in the absence of any shielding.

## 4.3 Radiation Hardness

Section 3 reports that the HM signal remains resistant to high dose rates with a net relative decrease of only 0.025%/kGy over a total dose of 216 kGy delivered in 15 minutes. This experiment delivered a highly accelerated dose with respect to any clinical application. This large dose delivered in such a short time period represents an acceleration factor of > $5\times10^4$ with respect to a nominal maximum clinical usage of 400 Gy/day, (i.e., 10 Gy/patient, 20 patients/day and a 2x safety factor). In our radiation test, not only was the time-averaged dose rate very high, but so was the instantaneous dose rate, at 8 GGy/s. A benchmark for electron FLASH instantaneous dose rate is 1 Gy in a 2 $\mu$s pulse, or 0.5 MGy/s. As a general rule, large acceleration factors often present unrealistically large worst-case scenarios, because they can produce multi-particle radiation damage mechanisms that under normal circumstances of reduced irradiation have a much lower probability of occurring. They also don't provide time for radiation damage recovery in air, which might be significant. The HM scintillator, could be expected to be irradiated at daily clinical doses for a year or more with less than 2.5% total signal loss under worst case conditions, without even allowing for any partial recovery from room temperature thermal annealing that occurs in inorganic scintillators over a time scale of days [44]. In reality, the signal loss after a year of FLASH-RT exposure is probably on the order of 1%, allowing for some level of scintillator recovery, the effect of enhanced radiation degradation caused by non-linear events resulting from the large acceleration factor and the fact that all of the dose was delivered in exactly the same spot.

Additionally, an external outrigger x-ray source (Moxtek Inc, Orem, UT) will be used to monitor the long-term stability. Periodic illumination of the scintillator produces an image that can be measured against a reference. Differential light yields between the reference and the current image are then used to generate a spatially dependent signal correction matrix. This matrix records the spatially dependent signal amplitude corrections registered by the calibration procedure. A measurable average global signal loss exceeding 1% would be used to indicate scintillator replacement.

## 4.4 Radiation Damage to Hardware

The radiation shielding also mitigates potential damage to the camera, while DAQ electronics will be shielded separately. In the context of an earlier ion beam monitor program, we have conducted limited radiation damage testing of an unshielded, sacrificial CMOS-based machine-vision camera type CamE. It was positioned adjacent to a FRIB beamline and passively exposed for 135 days, receiving a total (neutron) exposure of over 23,000 mSv, measured by an in-house monitor.

The before/after metrics were: The ADC gain and linearity of response to a light source independently measured by a photodiode, the RMS noise level, and number of dead pixels. None of these metrics registered statistically significant changes after the exposure. To translate this test to a clinical exposure we note that the dose (neutrons plus photons) delivered during proton beam therapy to healthy tissue distal (> 5 cm) from the target tumor is less than 1 mSv/Gy [45]. Additionally, the camera shielding reduces photon exposure five-fold, and an unknown amount for neutrons. The above-described test corresponds to 23-100 kGy therapeutic dose. The exposure to the DAQ, several meters removed from the beam and fully shielded will be considerably less. Further accelerated tests are planned where both camera and DAQ are directly irradiated, and where the single event upset error rate is the DAQ degradation metric.

## 4.5 Absolute Calibration

The FBSM with HM scintillator acquires the total signal integrated over a measured beam spot area and has been demonstrated to be linear from conventional RT through FLASH dose rates. Therefore, the device can be calibrated to absolute dose against a Farmer-type ion chamber at conventional dose rates in the calibration geometry specified by AAPM Task Group 51 for photon and electron beams, and by IAEA TRS-398 for proton beams. Gafchromic film is used for an independent secondary dosimetric verification and to verify geometrical accuracy. We note also that the light yield response of scintillators with respect to the linear energy transfer (LET) may generally be subject to Birks quenching [46] that empirically relates the light yield per unit pathlength dL/dx to the LET dE/dx as: $\frac{dL}{dx} = S \frac{dE}{dx} \frac{1}{1 + K_B \frac{dE}{dx}}$ where $K_B$ is the Birks constant and S is the photon conversion efficiency. While this parameter is unmeasured for HM, studies indicate that at the low LETs relevant to electron and proton energies in RT, light yield scales linearly with LET [47]. Nonetheless the calibration protocol for the FBSM would necessarily be conducted for monochromatic energy specific dose deliveries. Additionally for proton beams, as a transmission monitor it is not intended to be used in the Bragg peak where the LET changes are the greatest.

## 4.6 Other Scintillator Imaging Systems

As noted earlier, other groups are investigating scintillator-based imaging system's potential for FLASH applications. A high-speed proton beam imager [22] uses a large area scintillator, but it is very thick (5 cm) and thus suffers from excessive energy loss. A scintillation imaging system using a BaFBr:Eu scintillator was staged in a scanning pencil proton beam [24] and was characterized at continuous mode temporal frame rates of 10 ms/frame (100 Hz) and with 1 mm$^2$ spatial resolution. This implies that for a nominal beam dwell time of 1 ms (as expected e.g., in a FAST-01 clinical trial) several full multi-Gray doses are acquired into a single frame. Such a system would be incapable of generating an interrupt signal before the full localized dose is delivered. Another related system similarly employs a CMOS camera viewing a scintillator sheet [27]. It was operated at 1 KHz frame rate over a 10 cm × 10 cm field. Unlike the compact, enclosed, light-tight and high spatial resolution system (37 μm) described here, the camera viewed the scintillator from a large distance of ~ 2m which admits ambient light backgrounds, and possibly limits the spatial resolution to 1 mm in each coordinate. This is a promising approach, but a number of questions remain open regarding adaptability to clinical use with a patient, radiation hardness, image quality of the scintillator, mass profile and effective area that can be covered using a remote external camera. Additionally, the capability to process data rapidly for effective clinical proton or electron/photon FLASH use remains to be demonstrated since all data analysis was conducted offline and for proton beams only. Also, the real-time performance at >10 kHz frame rates, and radiation hardness of the system at FLASH-compatible ultra-high dose rates was not established.

## 4.7 Towards a Next Generation Clinical FBSM

The prototype FBSM results presented here have informed the design of the second generation currently under development. Several upgrades are included: The sensitive area will increase to $15 \times 23$ cm$^2$, sufficient for most if not all anticipated pre-clinical and clinical trials. The scintillator region will be viewed by two cameras (type CamE or CamP) mounted on opposite sides of the scintillator box using similar folded optics as in the prototype. The field of view of each camera covers a large section of scintillator and can provide an enhancement to spatial resolution. Importantly, the two-camera system also provides redundancy in the central region. The FPGA-based DAQ is already enabled to receive multiple high bandwidth camera streams that can be processed in parallel and therefore without introducing significantly longer timing latencies.

For pre-clinical (animal) trials and ultimately clinical applications, two critical development efforts are being pursued. First is the incorporation of trigger and timing signals necessary to synchronize camera frames. For scanning proton beams, which are done in a triggerless mode, this synchronization signal is in the form of a start-of-beam and end-of-scan signals. The start of beam initiates free-running data acquisition at 20 kHz until arrival of the end-of-scan pulse. For FLASH electron beams the FBSM operates in triggered mode wherein a beam pulse is used to initiate individual camera frames. This mode of operation has already been demonstrated at NDRL, as described earlier. Secondly, the clinical program of the planned dose in (X,Y) beam coordinates is being implemented into the FPGA firmware in the form of a look-up-table (LUT). The data in the LUT are transformed first into the camera coordinate system using a homography transform. The dose is also transformed into the equivalent charge density using the calibration protocols outlined earlier. The LUT is initialized before the start of data acquisition. The LUT is then referenced after each frame analysis is done. In this step, the measured location specific dose is compared to the planned clinical dose stored in the LUT for the same location. Any deviation issued by the comparator beyond a pre-set threshold triggers an interrupt signal sent to the beam interlock.

# 5 CONCLUSIONS

The FBSM described in this paper is designed to provide real-time precise spatially dependent dosimetry of beam profiles without significantly degrading the beam quality over a large area. A prototype device has been staged in pulsed electron particle beams at ultra-high dose rates consistent with FLASH RT applications. The HM scintillator was shown to be radiation hard with ≤ 0.025%/kGy degradation measured under highly accelerated, worst case conditions. True 2D transverse beam profiling, high spatial resolution (37 um), and a linear dose response was demonstrated using this novel scintillator. Combined data sets from FRIB and NDRL show that the HM scintillator responds linearly from single heavy ion particles to nearly 240 Gy/s dose rates, the highest rate tested. In a clinical electron beam test, the spatial reproduction of the beam profile was comparable to radiochromic films. The HM scintillator is radiation tolerant under nominal and accelerated clinical exposures. To establish fast real-time operation, a beta version of the continuous proton beam compatible firmware that runs on an FPGA using a camera type CamP was tested in the laboratory using the prototype FBSM. Operating at an ultrafast frame rate of 20,000 fps (50 μs/frame), we demonstrated fast beam analysis within < 1 μs.

## ACKNOWLEDGMENTS

This research has been supported by SBIR Phase-II awards to Integrated Sensors, LLC from both the NIH National Cancer Institute (Award: 1R44CA257178-01A1) and the DOE Office of Science and Office of Nuclear Physics (Award: DE-SC0019597).

*Disclaimer: This report was prepared as an account of work sponsored by an agency of the United States Government. Neither the United States Government nor any agency thereof, nor any of their employees, makes any warranty, express or implied, or assumes any legal liability or responsibility for the accuracy, completeness, or usefulness of any information, apparatus, product, or process disclosed, or represents that its use would not infringe privately owned rights. Reference herein to any specific commercial product, process, or service by trade name, trademark, manufacturer, or otherwise does not necessarily constitute or imply its endorsement, recommendation, or favoring by the United States Government or any agency thereof. The views and opinions of authors expressed herein do not necessarily state or reflect those of the United States Government or any agency thereof.*

## CONFLICT OF INTEREST

Co-author Dr. Peter Friedman, President and CEO of Integrated Sensors, LLC, owns intellectual property rights to the innovations described in this paper.